\pdfoutput=1
\documentclass[aps,amsmath,amssymb,reprint,showkeys,superscriptaddress]{revtex4-1}
	
\usepackage{charter,graphicx}
\usepackage{float}
\usepackage[version=4]{mhchem}
\usepackage{bm}

\usepackage[utf8]{inputenc}
\usepackage[T1]{fontenc}
\begin{document}

\title{Controlling Dzyaloshinskii-Moriya interactions in the skyrmion host candidates FePd$_{1-x}$Pt$_x$Mo$_3$N}

\author{Linus~Kautzsch}
\affiliation{Materials Research Laboratory, University of California, Santa Barbara, California 93106, USA}
\affiliation{Max Planck Institute for the Chemical Physics of Solids, D-01187 Dresden, Germany}

\author{Joshua~D.~Bocarsly}
\email{jdbocarsly@mrl.ucsb.edu}
\affiliation{Materials Research Laboratory, University of California, Santa Barbara, California 93106, USA}
\affiliation{Materials Department, University of California, Santa Barbara, California 93106, USA}

\author{Claudia~Felser}
\affiliation{Max Planck Institute for the Chemical Physics of Solids, D-01187 Dresden, Germany}

\author{Stephen~D.~Wilson}
\affiliation{Materials Research Laboratory, University of California, Santa Barbara, California 93106, USA}
\affiliation{Materials Department, University of California, Santa Barbara, California 93106, USA}

\author{Ram~Seshadri}
\affiliation{Materials Research Laboratory, University of California, Santa Barbara, California 93106, USA}
\affiliation{Materials Department, University of California, Santa Barbara, California 93106, USA}
\affiliation{Department of Chemistry and Biochemistry, University of California\\ 
Santa Barbara, California 93106, USA}

\date{\today}

\begin{abstract}
Ferromagnets crystallizing in structures described by chiral cubic space groups, including compounds with the 
B20 or $\beta$-Mn structures, are known to host long-period chiral spin textures such as skyrmion lattices. 
These spin textures are stabilized by a competition between ferromagnetic exchange and antisymmetric 
Dyzaloshinskii-Moriya (DM) exchange, which is enhanced by the spin-orbit coupling associated with high-atomic-number
elements. For real-world application, it is desirable to find materials that can host compact skyrmion lattices
at readily accessible temperatures. Here, we report on the crystal chemistry and magnetic phase diagrams of 
a family of compounds with the filled $\beta$-Mn structure, FePd$_{1-x}$Pt$_x$Mo$_3$N with $T_C$ 
ranging from 175\,K to 240\,K. DC and AC magnetization measurements reveal magnetic phase diagrams consistent 
with the formation of a skyrmion pocket just below $T_C$. The magnitudes of ferromagnetic and DM 
exchanges are determined from the phase diagrams, demonstrating that the introduction of increasing amounts of
Pt can be used to increase spin-orbit coupling in order to control the expected skyrmion lattice parameter
between 140\,nm and 65\,nm while simultaneously increasing $T_C$.
\end{abstract}


\maketitle


\section{Introduction}

Magnetic skyrmions are topologically protected vortex spin textures which act like single particles and may be 
manipulated as magnetic bits. These quasiparticle spin textures are promising candidates for magnetic 
racetrack-memory and other low-power, high-density spintronic devices \cite{Felser13, Fert13}. Magnetic
skyrmions were first observed in B20 compounds (space group $P2_13$) MnSi \cite{Muehlbauer09} and 
Fe$_{0.5}$Co$_{0.5}$Si \cite{Munzer10,Yu10}. In this structure type, breaking of inversion symmetry generates
long-range Dzyaloshinskii-Moriya (DM) exchange driven by spin-orbit coupling 
\cite{Bogdanov94,Dzyaloshinksii58,Moriya60}. Competition between this antisymmetric exchange and the usual 
ferromagnetic exchange leads to stabilization of long-period chiral spin structures, including helimagnetic, 
conical, and skyrmion lattice states.

It is of particular interest to develop new materials that exhibit skyrmions near room temperature with 
small diameters of the skyrmion vortex. Most known bulk hosts form stable skyrmion lattices only in a 
small pocket in the temperature and field phase diagram just below the Curie temperature: around 30\,K 
for both MnSi and Fe$_{0.5}$Co$_{0.5}$Si \cite{Bauer16,Bauer12}. In order to stabilize skyrmions at 
higher temperatures, one must look to chiral magnets with high Curie temperatures (\emph{i.e.} strong 
ferromagnetic exchange). In the B20 compound FeGe, which has $T_C$\,=\,278\,K, skyrmions have 
been observed near room temperature \cite{Yu2011}. More recently, skyrmion lattices at temperatures 
at and above room temperature have been studied in the compounds Co$_x$Zn$_y$Mn$_z$ ($x+y+z=20$) 
with the $\beta$-manganese crystal structure (space group $P4_132$ or $P4_332$) \cite{Tokunaga15}. As
in the B20 compounds, these materials exhibit a stable skyrmion pocket just below $T_C$, which 
is tunable between about 150\,K and 475\,K. Remarkably, by field-cooling through this pocket one may 
also observe a long-lived metastable skyrmion lattice which exists in a broad range of temperatures 
and fields \cite{Karube17,Karube18}. This phenomenon is believed to be caused by pinning of the skyrmions 
by the atomic site disorder \cite{Munzer10, Ritz13, Karube17, Karube18, Bocarsly19}. 

The skyrmion lattice parameter is controlled by a balance between the ferromagnetic exchange, which supports 
parallel alignment of spins, and the DM exchange, which supports rotation of near-neighbor spins with respect 
to one-another. The lattice parameter is therefore approximately given by $2 \pi A/D$ \cite{Kataoka87}, 
where $A$ is the ferromagnetic exchange stiffness and $D$ is the DM exchange 
density \cite{Bogdanov94,Heide09,Roessler11}. 
In principle, one may therefore decrease the skyrmion size by introducing heavy atoms into a skyrmion host 
to increase spin-orbit coupling and increase $D$. This approach has been successful in thin film 
heterostructures, where adding a layer of a heavy element such as Pt next to a magnetic layer can 
tune SOC \cite{Soumyanarayanan17}. However, in bulk materials this approach has not been straightforward to 
implement because chemical substitutions of heavy elements tend to have the dominant effect of weakening 
ferromagnetic exchange stiffness $A$. This may result in smaller skyrmions, but at the cost of dramatically lowering 
the temperature at which the skyrmion lattice phase is observed. For example, in Mn$_{1-x}$Ir$_{x}$Si, 
the skyrmion lattice parameter decreases by a factor of two for $x$\,=\,0.06, but the $T_C$ decreases by a 
factor of three \cite{Dhital17}. In Fe$_{1-x}$Mn$_x$Ge a composition change of $x$\,=\,0.2 decreases the 
skyrmion lattice parameter by a factor of four, while $T_C$ decreases by roughly 20\,\% \cite{Shibata13}.

Battle and co-workers \cite{Prior03,Prior04,Prior05,Battle07} have reported the formation of 
ferromagnetic nitrides in the filled-$\beta$-Mn structure, Fe$_{2-x}M_x$Mo$_3$N ($M$ = Co, Ni, Rh, Pd, Pt) 
with Curie temperatures up to 225\,K ($M$\,=\,Pt). This structure type, shown in Fig.~\ref{fig:diff_struc_dsc}~(a), 
is simply the $\beta$-Mn structure type with additional interstitial B, C, or N atoms. As these compounds 
have the chiral spacegroups of $\beta$-Mn ($P4_132$ or $P4_332$) and show ferromagnetic behavior, they 
are good candidates for skyrmion formation. Indeed, skyrmions in the filled $\beta$-manganese compound 
Fe$_x$Co$_{1.5-x}$Rh$_{0.5}$Mo$_3$N at temperatures around 100\,K have been reported using Lorentz transmission 
electron microscopy and magnetic measurements \cite{Li16}.

Here, we report the preparation of FePd$_{1-x}$Pt$_x$Mo$_3$N as a series with 5 compounds 
($x$\,=\,0, 0.25, 0.5, 0.75, 1) and the characterization of their structural and magnetic properties. 
Magnetic phase diagrams are determined, revealing an A-phase pocket just below the Curie temperatures, which 
is expected to host a skyrmion lattice. An analysis of these phase diagrams shows that the introduction of 
Pt atoms, associated with large spin-orbit coupling, leads to an increase in $D$ while leaving $A$ largely
unchanged. Therefore, this substitution dramatically lowers the expected skyrmion lattice parameter while actually 
enhancing $T_C$.  This study demonstrates how the introduction of heavy atoms can be used to control chiral magnetism
in a bulk material, in a system where such a substitution does not weaken ferromagnetic exchange.

\section{Methods}

\subsection{Sample preparation}
 
Polycrystalline samples of FePd$_{1-x}$Pt$_x$Mo$_3$N ($x=0$, 0.25, 0.5, 0.75, 1) were prepared by direct 
nitridation of the metals \cite{Prior19}. The appropriate transition metals were intimately ground 
in stoichiometric mixtures, pelletized and heated under a gas flow of 5\% H$_2$ in N$_2$ 
(flow rate $\approx$\,7\,dm$^3$\,min.$^{-1}$) for 36\,h at 975\,$^{\circ}$C. The samples were then re-ground 
and re-pelletized and heated once again at the same conditions for 24\,h. In both steps of the process,
the furnaces were purged with the reaction gas prior to heating and the samples were cooled down to room 
temperature under gas flow. Samples with a total mass of 500\,mg were prepared. Reagents employed: Fe (Sigma-Aldrich) 99.9\%, Pd (Sigma-Aldrich) 99.9\%, 
Pt (Alfa Aesar) 99.95\%, Mo (Sigma-Aldrich) 99.9\%. This thermal treatment procedure gave well-crystalized 
phase-pure samples according to laboratory X-ray diffraction, but in order to further improve on the 
sample crystallinity and density, the samples were again re-ground and re-pelletized and annealed in 
amorphous silica tubes filled with 1\,atm N$_2$ for 48\,h. In order to contain this pressure and temperature 
of N$_2$ gas safely, thick-walled silica tubes were chosen (outer diameter: 15.8\,mm,  wall thickness: 2.8\,mm, 
length: $\approx$7\,cm). The samples were annealed at the following temperatures: $x=0$ at 1000$^{\circ}$C, 
$x=0.25$ at 1050$^{\circ}$C, $x=0.5$ at 1100$^{\circ}$C, $x=0.75$ at 1150$^{\circ}$C and $x=1$ at 1200$^{\circ}$C. 
These annealing temperatures were chosen to be about 50\,$^{\circ}$C below the decomposition temperature of 
each sample, as determined by Differential Scanning Calorimetry.

\subsection{Characterization} 

The progress after each step of the synthesis was verified by X-ray powder diffraction using a Panalytical 
Empyrean powder diffractometer operating with Cu-K$_{\alpha}$ radiation in Bragg-Brentano geometry. Differential scanning 
calorimetry was performed using a Netzsch DSC 404 C Pegasus Thermal analyzer equipped with a high temperature 
furnace (up to 1500~$^{\circ}$C). Samples with a mass of about 20\,mg were measured in \ce{Al2O3} crucibles 
in flowing \ce{N2} gas. After successful preparation, high-resolution synchrotron powder diffraction data 
was collected for all samples in the series at beamline 11-BM at the Advanced Photon Source (APS), Argonne 
National Laboratory, using an average wavelength of 0.457838\,\AA. The patterns were collected at a temperature
of 295\,K. Elemental analysis was performed on powder samples on Kapton tape using a Rigaku ZSX Primus IV X-ray fluorescence (XRF) spectrometer.

Rietveld refinement of the crystal structures was carried out using TOPAS academic. The 
synchrotron patterns were fit with a Stephens strain model \cite{Stephens99} (two parameters), and a standard 
Gaussian size broadening term (one parameter). A fixed axial divergence asymmetry correction based on the 
instrument geometry was used to handle diffractometer peak asymmetry. Absorption of the X-rays by the capillary 
samples was corrected for according to the chemical compositions and capillary diameter (0.8\,mm) by refining 
the packing density of the samples to values between 30\% and 40\%. Unit cell parameters, atom positions as well 
as the elemental compositions were refined, assuming the atomic sites were each fully occupied. For the samples 
containing both Pd and Pt, the compositions were additionally restricted to stoichiometries consistent with the 
nominal Pd/Pt ratio, \emph{e.g.} for $x=0.25$: $x_{\mathrm{Pd}} = 0.25 \cdot (1-x_{\mathrm{Fe}})$ and 
$x_{\mathrm{Pt}} = 0.75 \cdot (1-x_{\mathrm{Fe}})$. Isotropic displacement parameters ($B_{\rm{iso}}$) were refined 
for the two metal-filled sites in the unit cell. The value for the nitrogen-filled 4$a$ site was fixed to a 
refined overall $B_{\rm{iso}}$.

DC and AC magnetic measurements between 2\,K and 400\,K were obtained on a Quantum Design MPMS3 SQUID magnetometer. 
Pieces with masses less than 1\,mg were measured in polypropylene capsules. First, DC magnetization as a function 
of temperature between 2\,K and 400\,K was collected under a constant field of 0.02\,T while sweeping temperature 
at a rate of 7\,K\,min$^{-1}$. Five-branch hysteresis loops between $-$7\,T and $7$\,T were then collected 
at 2\,K at a rate of 10\,mT\,s$^{-1}$. Detailed magnetic phase diagrams for the $x=0$, 0.5, and 1 samples were 
constructed using DC magnetization and AC susceptibility measurements as a function of field at closely spaced 
temperatures near the magnetic transition. The AC susceptibility measurements were performed every 2\,K at 
temperatures below the Curie transition using excitation frequencies of 10\,Hz and amplitudes of 0.5\,mT. 
The susceptibility was measured every 0.7\,mT between 0 and 25\,mT and every 5\,mT between 25\,mT and 100\,mT. 
The DC magnetization was measured every 1\,K while sweeping at a rate of 0.2\,mT\,s$^{-1}$ between 0 and 100\,mT, 
and collecting a data point every 1 second. The field derivatives of the AC and DC measurements were calculated 
using Tikhonov regularization \cite{Stickel10}.

Additionally, the magnetoentropic behavior of \ce{FePtMo3N} was characterized using the technique described 
in Ref.~\onlinecite{Bocarsly18}. DC magnetization versus temperature sweeps at closely spaced fields were 
performed around the magnetic transition temperature of \ce{FePtMo3N} to obtain magnetocaloric information. 
The sweeps were performed between 210\,K and 240\,K at a rate of 4\,K\,min$^{-1}$ with a spacing of 2\,mT 
between 2 and 30\,mT and a spacing of 5\,mT between 30\,mT and 80\,mT. The temperature derivatives of 
these sweeps were then calculated using Tikhonov regularization as well to give information about how the 
entropy changes with applied field at any given temperature and field, according to the Maxwell relation:

\begin{equation}
\left(\frac{\partial M}{\partial T}\right)_H = \left(\frac{\partial S}{\partial H}\right)_T
\label{eq:maxwell}
\end{equation}

\noindent where $M$, $T$, $S$, and $H$ are the magnetization, temperature, entropy, and field, respectively. 
This quantity can then be integrated with respect to field, to give the isothermal magnetic entropy 
change $\Delta S_M(T,H)$, which is the entropy difference between the sample at a magnetic field $H$ and at 
zero field:

\begin{equation}
\Delta S_M(T,H) = \int_0^H \left(\frac{\partial M}{\partial T}\right)_{H'} \mathrm{d} H'
\label{eq:DSm}
\end{equation}

\section{Results and discussion}

\subsection{Crystal structure}

\begin{figure}[ht]
	\centering
	\includegraphics[height=14cm]{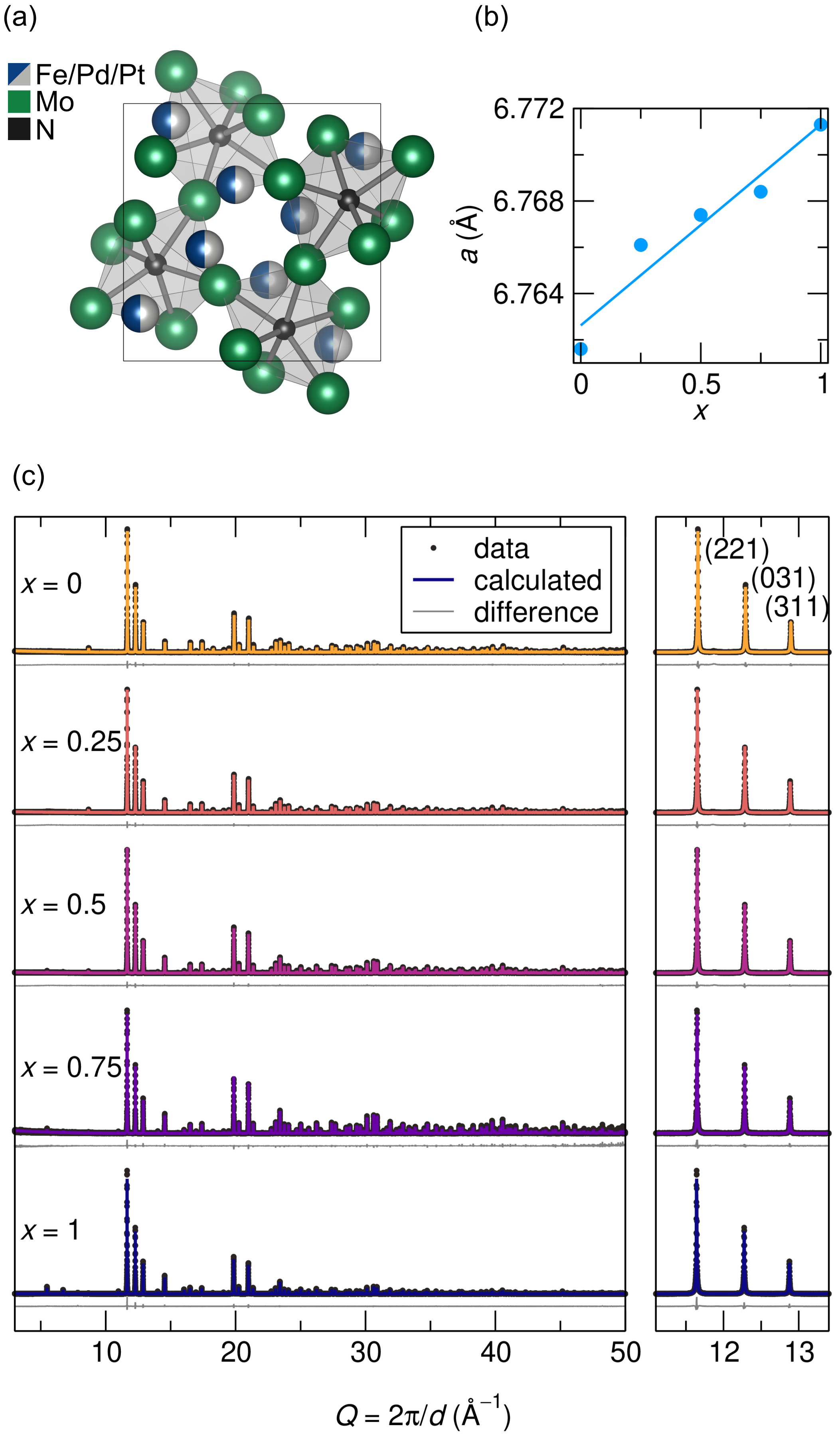}
	\caption{(a) Crystal structure of the cubic series FePd$_{1-x}$Pt$_x$Mo$_3$N (space-group $P4_132$), 
shown projected down one of the cubic axes. The origin of the cell has been translated by 0.25 of a unit 
cell such that the chiral fourfold screw axis ($4_1$) is centered in the cell. (b) Evolution of the lattice 
parameter $a$ as a function of $x$ (error bars are smaller than the points). (c) Synchrotron diffraction pattern shown with Rietveld fits 
to the $P4_132$ $\beta$-Mn structure.}
	\label{fig:diff_struc_dsc}
\end{figure}

\begin{table}[!hbt]
\setlength{\tabcolsep}{4pt}
\small
\caption{Results of Rietveld refinement. Lattice parameter $a$; $x$ coordinate of the 8$c$ position; 
$B_{\rm{iso}}$ value of the 8$c$ position; $y$ coordinate of the 12$d$ position; 
$B_{\rm{iso}}$ value of the 12$d$ position; fraction $x$ of Fe, Pd and Pt on the 8$c$ position; 
$R$-factors $R_{\mathrm{wp}}$ and $R_{\mathrm{exp}}$.}
\begin{tabular}{lrrrrr}
\toprule
$x$       & 0          & 0.25       & 0.5        & 0.75       & 1          \\
\colrule
$a$ (\AA) & 6.762      & 6.766      & 6.767      & 6.768      & 6.771      \\
8$c$ $x$  & 0.0654 & 0.0656 & 0.0654 & 0.0653 & 0.0650 \\
8$c$ $B_{\rm{iso}}$ & 0.91(2)    & 0.76(2)    & 0.68(1)    & 0.60(2)    & 0.90(1)  \\
12$d$ $y$ & 0.2046 & 0.2045 & 0.2043 & 0.2042 & 0.2041 \\
12$d$  $B_{\rm{iso}}$ & 0.79(1)    & 0.68(1)    & 0.63(1)    & 0.47(2)    & 0.77(7)    \\
$x_{\rm{Fe}}$   & 0.500(8)   & 0.500(5)   & 0.503(3)   & 0.501(5)   & 0.491(2)   \\
$x_{\rm{Pd}}$   & 0.500(8)   & 0.375(3)   & 0.248(1)   & 0.125(1)   & \multicolumn{1}{c}{--}           \\
$x_{\rm{Pt}}$    & \multicolumn{1}{c}{--}         & 0.125(1)   & 0.248(1)   & 0.374(4)   & 0.509(2)   \\
$R_{\rm{wp}}$  & 21.02      & 18.95      & 16.56      & 29.04      & 15.93      \\
$R_{\rm{exp}}$ & 17.04      & 16.13      & 13.36      & 30.41      & 10.79     \\
\botrule
\end{tabular}
\label{tab:refin}
\end{table}

Rietveld refinements of synchrotron powder diffraction patterns for the compounds in the series 
FePd$_{1-x}$Pt$_x$Mo$_3$N are shown in Fig. \ref{fig:diff_struc_dsc}(c). The reported $\beta$-Mn structure 
($P4_132$) \cite{Prior19} fits the obtained diffraction patterns very well. Sharp diffraction peaks are observed, 
and no peaks violating this model or suggesting the presence of secondary phases are observed. The 8$c$ 
site is occupied by a combination of Fe, Pd, and Pt.  Mo fully occupies the 12$d$ site, forming distorted 
corner-sharing Mo$_6$N octahedra with the interstitial nitrogen atoms on the 4$a$ site, as shown in 
Fig.~\ref{fig:diff_struc_dsc}(a). The lattice parameter of the unit cell is found to increase with rising 
Pt content Fig.~\ref{fig:diff_struc_dsc}(b). The refined compositions are in good agreement with the results from the XRF measurements (Tab. S1). Details of the refinements are presented in table~\ref{tab:refin}. 

Differential scanning calorimetry (DSC) was employed to characterize the thermal stability of the compounds 
after the two gas nitridation steps. Upon heating under 1\,atm. of \ce{N2} gas all samples show an endothermal 
signal somewhere between 1100$^\circ$C ($x$\,=\,0) and 1250$^\circ$C ($x$\,=\,1) (Fig. S1), which corresponds 
to decomposition of the samples \emph{via} loss of nitrogen. Upon subsequent cooling, the nitrogen is 
not reabsorbed, and the resulting sample is found to be a mix of metal binaries. When heating in Ar, the 
decomposition temperatures were found to be roughly 100\,K lower, and so the annealing step for the sample 
preparation was performed under nitrogen pressure, using temperatures chosen to be about about 50\,K below 
the observed decomposition temperature of each sample. This annealing step is found to substantially improve 
sample crystallinity, as evidenced by narrow diffraction peak widths and the sharpness of the magnetic transitions. 
Because the $x$\,=\,1 sample has the highest decomposition temperature, it could be annealed at the highest 
temperature and consequently has the clearest magnetic features.

\subsection{Magnetic properties}

\begin{figure}[ht]
	\centering
	\includegraphics[height=8cm]{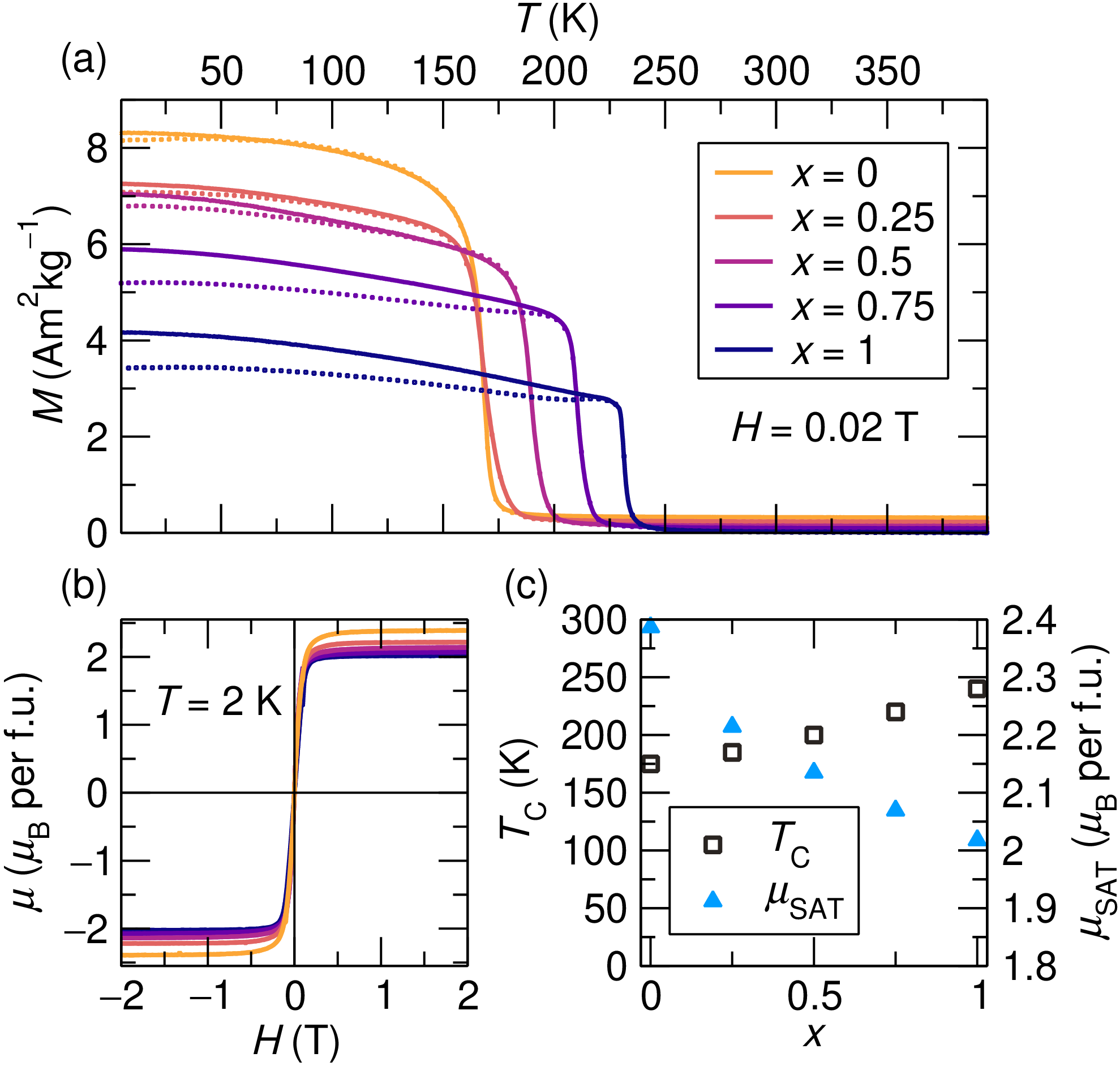}
	\caption{(a) Magnetization as a function of temperature collected under an applied magnetic field 
of $H$ = 0.02\,T, collected while warming after either cooling under zero field (dotted line) or cooling 
under a field (solid line). (b) Magnetization as a function of the applied magnetic field at $T$ = 2\,K. 
(c)  $T_C$ and $\mu_\mathrm{sat}$ as a function of $x$.}
	\label{fig:zfc-mh}
\end{figure}

All compounds of the series show soft ferromagnetic behavior and a saturated magnetic moment between 
2.0\,$\mu_{B}$ and 2.4\,$\mu_{B}$ \textit{per} Fe atom at $T$ = 2\,K [Fig.~\ref{fig:zfc-mh} (b), Table~\ref{tab:magn}]. The 
magnetization saturates at approximately $H$ = 0.4\,T. This behavior is consistent with other reports on this 
materials class \cite{Prior04,Prior05}. The zero field cooled warming and the field cooled warming curves in 
Fig.~\ref{fig:zfc-mh} (a) display a small amount of irreversibility below the Curie temperature, with the 
splitting between field-cooled and zero-field cooled curves increases with increasing Pt content. However, 
this irreversibility is too small to cause a visible hysteresis around the origin in the $M(H)$ hysteresis loops. 
The sharpness of the Curie transitions supports that the samples are of high quality.  $T_C$ was determined from the onset of the Curie transition.

\begin{table}[H]
\setlength{\tabcolsep}{6pt}
\renewcommand*{\arraystretch}{1.3}
\centering
\caption{Curie temperature $T_C$ and saturation magnetization $\mu_{\mathrm{sat}}$ \textit{per} Fe 
at $T$ = 2\,K and $H$ = 2\,T.}
\begin{tabular}{ccc}
\toprule
$x$ & $T_{\rm{C}}$ (K) & $\mu_{\rm{sat}}$ ($\mu_B$/Fe) \\ \colrule
0 & 175 & 2.39\\
0.25 & 185 & 2.21\\
0.5  & 200 & 2.14\\
0.75 & 220 & 2.07\\
1 & 240 & 2.02\\
\botrule
\end{tabular}
\label{tab:magn}
\end{table}

\subsection{Magnetic phase diagrams}

\begin{figure*}[htpb]
	\centering
	\includegraphics[height=6.2cm]{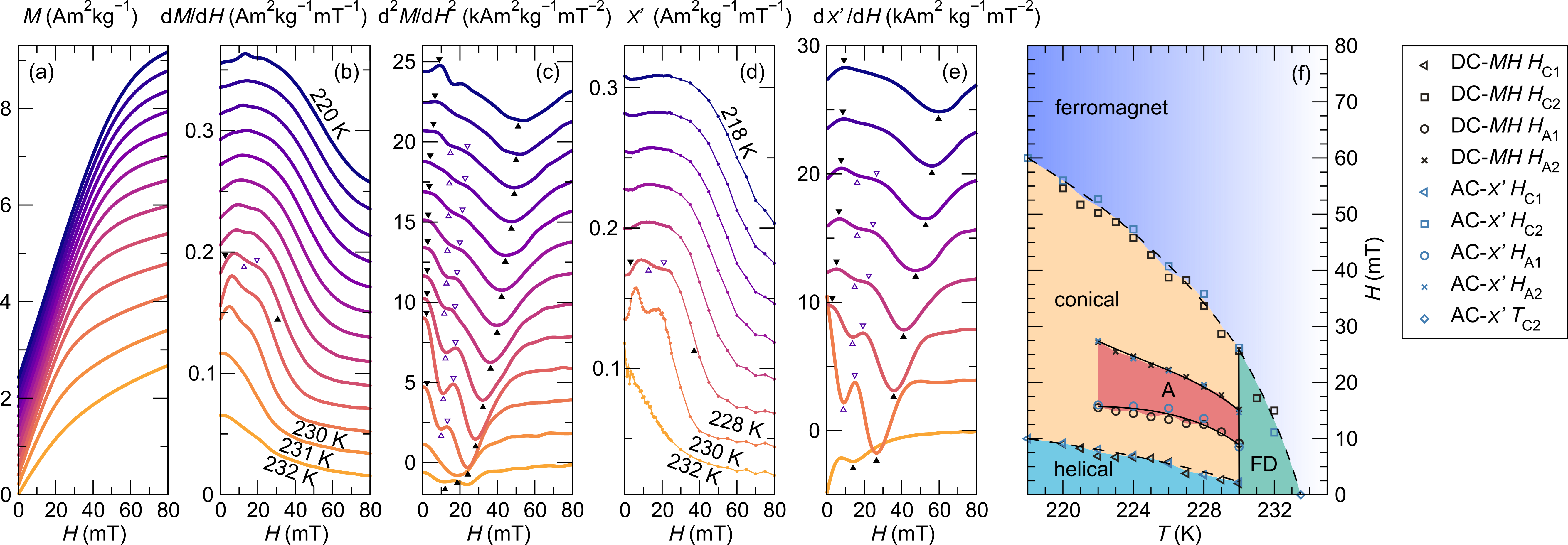}
	\caption{Magnetic characterization of FePtMo$_3$N (a-f). Magnetization versus magnetic field (a). 
For visual clarity, the curves are each offset by 0.2~Am$^2$kg$^{-1}$. (b) d$M$/d$H$ with an offset between 
the curves of 0.02~Am$^2$kg$^{-1}$mT$^{-1}$. (c) d$^2M$/d$H^2$ shifted by 2~kAm$^2$kg$^{-1}$mT$^{-2}$. 
(d) $\chi'$ with an offset of 0.01~Am$^2$kg$^{-1}$mT$^{-1}$ and (e) d$\chi'$/d$H$ shifted by 
4~kAm$^2$kg$^{-1}$mT$^{-2}$. d$^2M$/d$H^2$ curves (c) and d$\chi'$/d$H$ (e) below the magnetic transition 
temperature show features corresponding to magnetic phase transitions, which can be used to draw a phase 
diagram in (f). The proposed skyrmionic pocket (A) is shown in red. Solid lines represent first-order 
phase transitions. The dashed lines represent continuous transitions. FD: fluctuation disordered.}
	\label{fig:phasedia}
\end{figure*}

Magnetic phase diagrams for the $x$\,=\,0, 0.5, and 1 samples were constructed on the basis of DC magnetization and AC susceptibility
measurements as a function of field, collected at several temperatures near the Curie temperature. This process 
is shown in detail for \ce{FePtMo3N} ($x$\,=\,1) in Fig.~\ref{fig:phasedia}. Thirteen DC $M(H)$ curves and 
eight AC $\chi' (H)$ curves are shown. The AC susceptibility measurements are collected using a relatively 
low excitation frequency of 10\,Hz, and therefore should correspond approximately to the field derivative 
of the DC magnetization data $\mathrm{d}M/\mathrm{d}H$. Therefore, the DC and AC data are expected to provide the 
same information, although measurement and data analysis details may lead to minor differences. In practice, we find that the AC measurements serve as a high-sensitivity complement to the high-density data derived from DC magnetization.


Just below $T$ = 230\,K, a characteristic double-bump feature may be seen in both AC $\chi'(H)$ 
[Fig.~\ref{fig:phasedia}~(d)] and in $\mathrm{d}M/\mathrm{d}H$ calculated from the DC magnetization 
data [Fig.~\ref{fig:phasedia}~(b)]. This feature is characteristic of the phase evolution seen just 
below $T_C$ in chiral cubic magnets  \cite{Bauer16book, Bauer12, Adams12, Tokunaga15, Li16, Bocarsly18}, 
from a helical state at zero field (low susceptibility) to a canted conical state (high susceptibility), 
to a skyrmion lattice (low susceptibility), back to the canted conical state (high susceptibility) and 
finally to a field-polarized state (low susceptibility). The phase boundaries between these phases are found 
at the inflection points, which may conveniently be identified as local maxima and minima in the field 
derivative of the susceptibility, \emph{i.e.} AC $\mathrm{d}\chi/\mathrm{d}H$ [Fig.~\ref{fig:phasedia}~(e)] 
or DC $\mathrm{d}^2 M/ \mathrm{d}H^2$ [Fig.~\ref{fig:phasedia}~(c)]. Below about 222\,K, the transitions in 
and out of the skyrmion lattice phase (called $H_{\rm{A1}}$ and $H_{\rm{A2}}$ in Fig.~\ref{fig:phasedia}~(c,e)) 
disappear, but the transitions from helical to conical ($H_{\rm{C1}}$) and conical to field-polarized 
($H_{\rm{C2}}$) remain. Above $T$ = 230\,K, only $H_{\rm{C2}}$ is observed.

When these features are plotted on temperature and field axes, a magnetic
 phase diagram that is consistent with the phase diagram of other chiral cubic skyrmion host materials 
is obtained  [Fig.~\ref{fig:phasedia}~(f)]. This generically observed diagram features a helical ground 
 state that transforms into a canted conical phase and eventually a field-polarized ferromagnetic phase 
 upon application of field. Just below the Curie temperature, an ``$A$-phase'' pocket, which is expected to host a skyrmion 
 lattice structure, is found. In \ce{FePtMo3N}, this pocket is found between 10\,mT and 20\,mT and 222\,K and 
 230\,K. At 230\,K, a Brazovskii transition is found, which involves a line of first-order transitions from 
 the ordered chiral phases to a short-range ordered ``fluctuation-disordered'' phase  
 \cite{Janoschek13, Brazovskii75, Bauer16} that persists to 233.5\,K. The data points determining the phase 
 boundaries from DC and AC magnetic measurements are in good agreement.

\begin{figure}[ht]
	\centering
	\includegraphics[height=6.5cm]{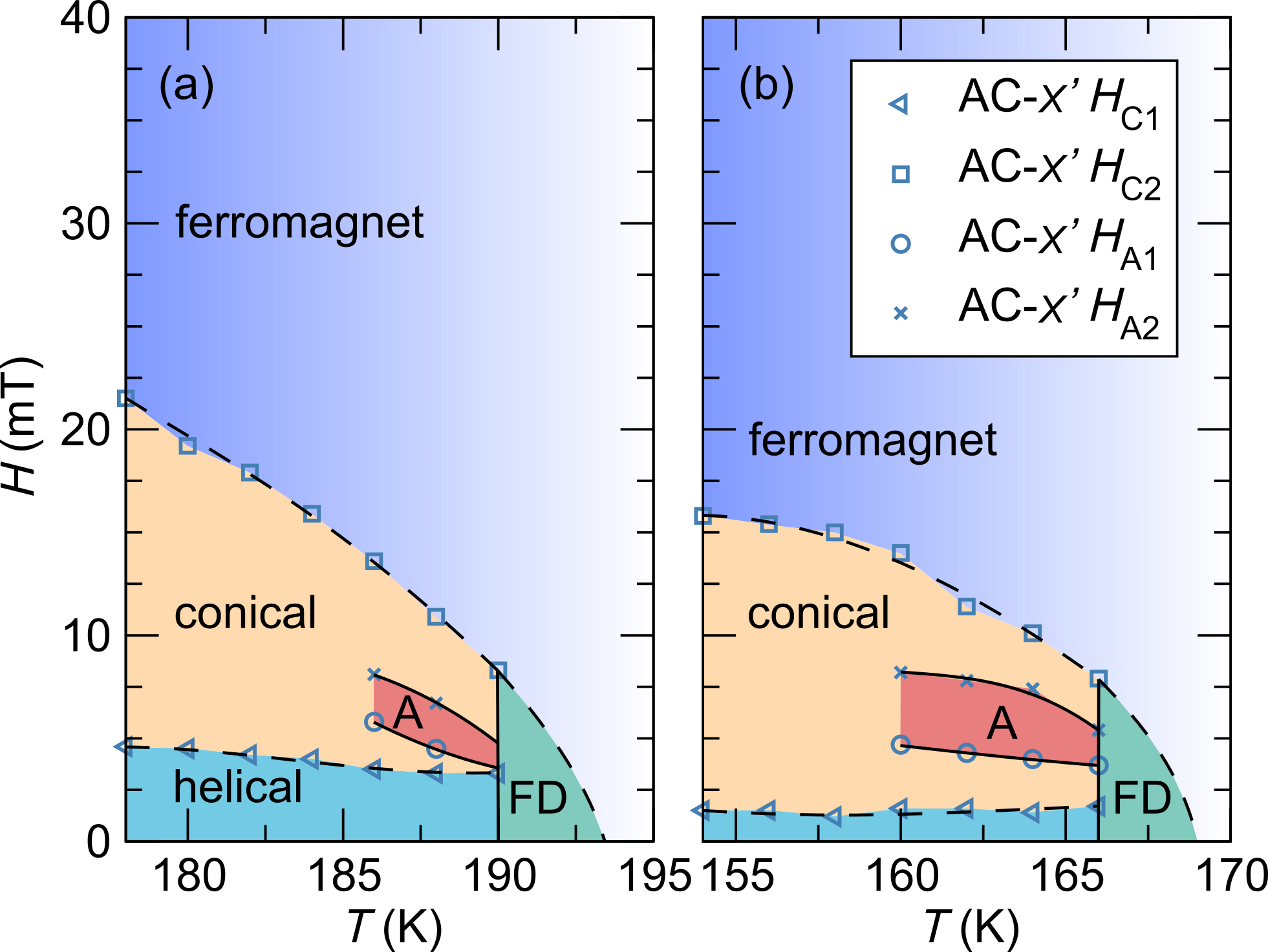}
	\caption{Magnetic characterization of (a) FePd$_{0.5}$Pt$_{0.5}$Mo$_3$N and (b) FePdMo$_3$N. d$\chi'$/d$H$ 
	has been used to construct the phase diagrams. The proposed skyrmionic pocket (A) is shown in red. 
	Solid lines represent first-order phase transitions. The dashed lines represent continuous transitions. 
	FD: fluctuation disordered.}
	\label{fig:ophase}
\end{figure}

Magnetic phase diagrams constructed from AC susceptibility measurements of the compounds with $x=0.5$ and $x=0$ 
can be found in Fig.~\ref{fig:ophase}. Both compounds exhibit signatures similar to FePtMo$_3$N. Skyrmion lattice 
pockets (A) are observed below the fluctuation disordered state (FD) starting at 190~K ($x=0.5$) and 166~K ($x=0$). 
The lower temperatures for the A regions correspond to the lower Curie transition temperatures of the compounds. 
Additionally, the A regions emerge at lower magnetic fields.

\begin{figure}[ht]
	\centering
	\includegraphics[height=7cm]{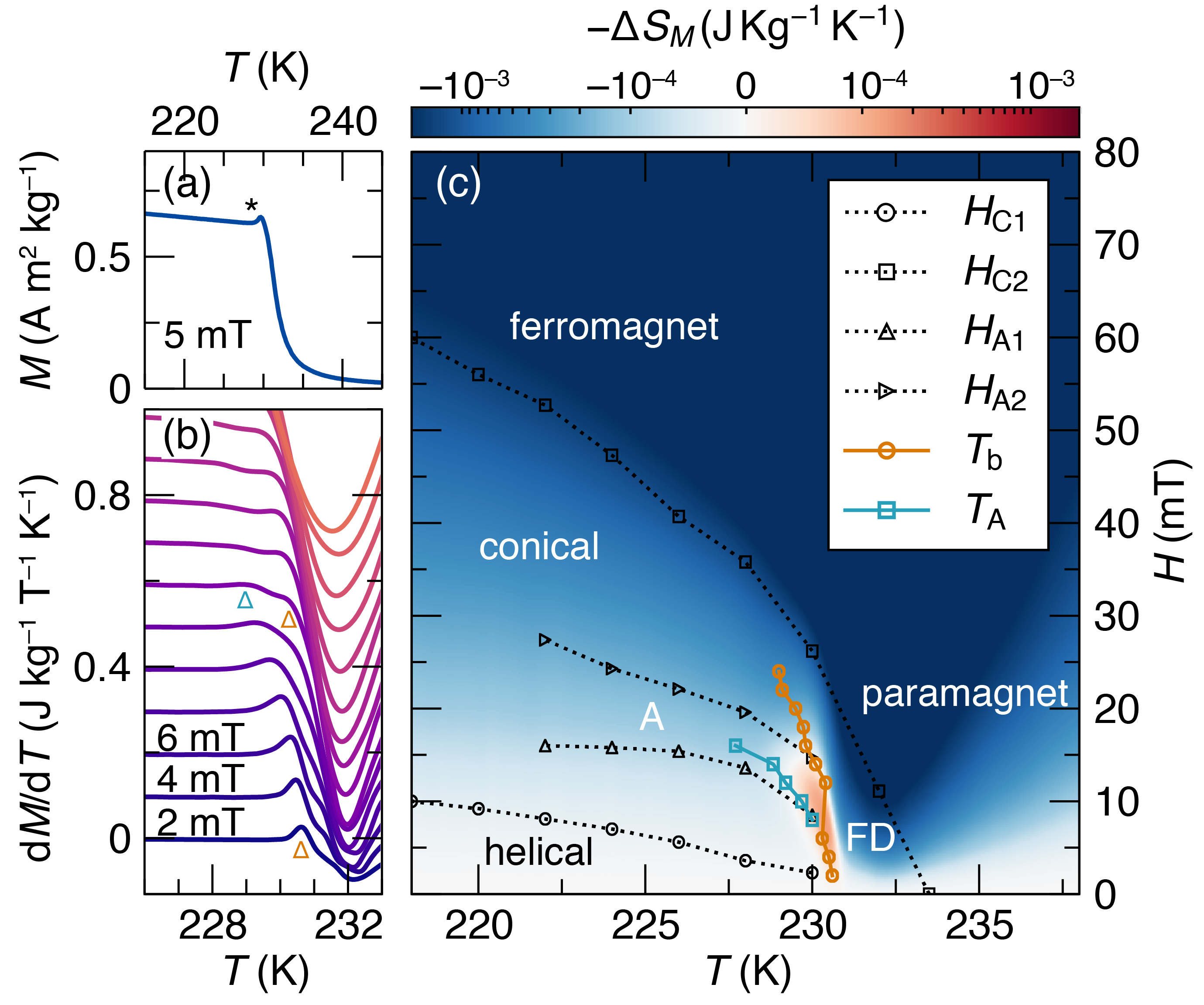}
	\caption{Magnetoentropic characterization of FePtMo$_3$N. (a) Magnetization \textit{vs.} temperature 
        shows precursor anomalies characteristic for skyrmion host materials, which are seen as features in 
        $\mathrm{d}M/\mathrm{d}T$ = $\mathrm{d}S/\mathrm{d}H$ (b). In (b), the orange triangles indicate the anomaly corresponding to the Brazovskii 
        transition, and the teal to the transition from the conical to A state. The curves are offset from each other 
        by 0.1\,J\,kg$^{-1}$\,K$^{-1}$\,T$^{-1}$ for clarity. (c) Map of the magnetocaloric effect ($\Delta S_M$) 
        with the features from $\mathrm{d}M/\mathrm{d}T$ indicated as orange and teal lines. The phase transition lines from AC 
        magnetic measurements are shown as black dotted lines.}
	\label{fig:magnetoentropy}
\end{figure}

Another way of looking at the magnetic phase diagram of magnetic skyrmion host is on the basis of magnetoentropic 
mapping, as was previously used to reveal bulk entropy changes associated with magnetic phase transitions in the 
chiral cubic skyrmion host FeGe \cite{Bocarsly18}. Using the fact that the temperature derivative of magnetization is equal to the 
field derivative of entropy (Eq.~\ref{eq:maxwell}), magnetization \emph{vs.} temperature measurements collected at 
many fields are used to create a map of the entropic response to field of the sample (Fig.~\ref{fig:magnetoentropy}). 
In $M(T)$ measurements of \ce{FePtMo3N} in low magnetic fields (\emph{e.g.} 5~mT), a characteristic cusp right 
below the Curie transition is apparent, shown with an asterisk in Fig.~\ref{fig:magnetoentropy}~(a). When the derivative of this 
$M(T)$ curve is taken, the cusp becomes a region of positive $\mathrm{d}S/\mathrm{d}H$ [Fig.~\ref{fig:magnetoentropy}~(b)]-- \emph{i.e.} 
application of a magnetic field \emph{increases} the entropy of the system. In a skyrmion host material, this counterintuitive 
phenomenon occurs both at the field-driven conical to skyrmion lattice transition ($H_{A1}$) and at the Brazovskii transition, 
where field drives the transformation from ordered phases to the fluctuation-disordered phase. For most other cases, 
application of a magnetic field decreases entropy by suppressing spin fluctuations. By tracking the positions of 
maxima in $\mathrm{d}S/\mathrm{d}H$, phase boundaries of these phase transitions can be determined.

Figure~\ref{fig:magnetoentropy}~(c) shows shows a map of the isothermal magnetic entropy changes $\Delta S_M(T,H)$ 
of \ce{FePtMo3N}, overlaid with the phase diagram determined using AC susceptibility. $\Delta S_M(T,H)$ is the integral with field
of $\mathrm{d}S/\mathrm{d}H$ [Eq.\,\ref{eq:DSm}]. In this map, red regions indicate 
increased entropy relative to the zero-field state, and blue indicates decreased entropy. The orange and teal points ($T_{\rm{b}}$ and $T_{\rm{A}}$)
represent the maxima in $\mathrm{d}S/\mathrm{d}H$, and generally agree with the position of the Brazovskii transition and the transition
 from the conical state to the A phase determined using AC susceptibility, although the feature associated with $T_{\rm{A}}$ is very subtle 
 and is not resolvable below 227\,K. We attribute this weak signal to the polycrystalline nature of the samples: because 
 this technique measures the bulk entropic response of the sample, grain boundaries and defects that do not host the 
 ordered skyrmion lattice phase will weaken the observed bulk entropy signal and contribute a net negative signal as 
 applied field decreases the spin disorder at these features. For comparison, the magnitude of the positive magnetoentropic 
 signals seen here is about 5 times less than were observed in single crystals of FeGe \cite{Bocarsly18}, but is 
 comparable to that seen in a polycrystalline sample of the skyrmion host \ce{Co8Zn9Mn3} \cite{Bocarsly19}. Despite this
  weakened signal, close to $T_C$ a pocket of increased entropy (the red area in the $\Delta S_M$ map) is resolvable, indicating 
  the formation of a phase with higher entropy than the helical ground state. This increased bulk entropy signal is 
  consistent with the formation of a bulk skyrmion lattice phase in the A phase region.

\subsection{Effect of spin-orbit coupling on magnetic exchange and Dzyaloshinskii-Moriya interaction}
While the phase diagrams for the \emph{x} = 0, 0.5, and 1 are qualitatively the same, the details of the temperatures 
and fields are expected to be controlled primarily by the ferromagnetic exchange stiffness ($A$) and 
Dzyaloshinskii-Moriya interaction densities ($D$). Furthermore, the helimagnetic wavelength and skyrmion lattice 
parameter are expected to be related to these constants according to $\lambda = 2\pi A/D$. Here, we use the 
determined magnetic phase diagrams to obtain estimates for $A$ and $D$ (and therefore skyrmion lattice parameter) 
as a function of Pt content $x$.

First, the observed Curie temperature is used to obtain $A$. A Monte-Carlo investigation of the phase diagram of the 
Heisenberg model for a cubic chiral magnet \cite{Buhrandt13} has determined that the observed magnetic transition 
temperature (specifically, the temperature of the Brazovskii transition) is:

\begin{equation}
T_C = \frac{J}{0.93~k_{\rm{B}}}
\end{equation}

\noindent Where $J$ is an effective exchange energy (units of eV) and $k_{\rm{B}}$ is the Boltzmann constant. 
In a cubic system, this exchange energy can be converted to an exchange density $A$ (units of eV\,nm$^{-1}$) using: 

\begin{equation}
A = \frac{J s^2}{a} n.
\end{equation}

\noindent where $a$ is the lattice parameter and $n$ coordination number (3 for the 8c site in the present 
structure) \cite{Chikazumi97}. $s$ is the spin value of the magnetic lattice, which in this case we take to 
be the average spin value of an atom on the 8c lattice, which is half occupied by magnetic Fe and half by 
nonmagnetic Pd/Pt). This value is determined using the 2\,K saturated magnetization.

The field at which the conical phase saturates to a ferromagnetic state ($H_{\mathrm{C2}}$) can then be used 
to determine $D$. If one ignores magnetocrystalline anisotropy, which is believed to be small due to the low helical to conical
transition fields observed, $H_{\mathrm{C2}}$ is expected to occur at:

\begin{equation}
\frac{\gamma \hbar}{V_0} \mu_0 H_{\rm{C2}} = \frac{D^2 \sigma(T)}{A}, 
\label{eq:hc2}
\end{equation}

\noindent where $V_0$, $\gamma$, $\hbar$ and $\mu_0$ are the volume of the formula unit cell, the gyromagnetic 
ratio, the reduced Planck constant and the vacuum permeability, respectively \cite{Kataoka86,Wilson14}. 
$\sigma(T)$ is the spin density at a given temperature. In this picture, the temperature-dependence of the ordered magnetic moment drives 
the temperature-dependence of $H_{\mathrm{C2}}$. This spin density is obtained from the spontaneous ordered 
magnetic moment, $M_{\rm{s}}(T)$ according to: 

\begin{equation}
\sigma(T) = \frac{M_{\rm{s}}(T) V_0}{\gamma \hbar}.
\label{eq:ms}
\end{equation}

\noindent
$M_{\rm{s}}$ at each temperature is deducted from the Arrot-Belov plot (Fig.~S2) \cite{Belov56,Arrot57,Kuzmin2010}. 
For each sample, six $M(H)$ sweeps with a spacing of 2~K were performed below the Curie temperature. 
Linear fits to the data with values of $H$/$M$ above 0.05\,T\,kg\,A$^{-1}$m$^{-2}$ were used to determine 
the $M_{\mathrm{s}}$ values from the $x$ intercepts. The parameter $D$ was obtained by Eq.~\ref{eq:hc2} 
using six values each for $H_{\mathrm{C2}}$ and $\sigma(T)$ below the Brazovskii transition. 
Table~\ref{tab:aver_fie} shows $J$, $s$, $A$, $D$ and the helical wavelength $\lambda$ for 
the compounds of the series. We note that the approximations made during this analysis, such as the conversion 
factor between $T_C$ and $J$ and neglecting the anisotropy terms may lead to errors in the obtained $J$ and 
$D$ parameters. In particular, neglecting anisotropy may cause a slight overestimation of $D$ and therefore
a small underestimation in $\lambda$. However, we do not expect that the overall trends, especially with regard to expected helical 
wavelength $\lambda$ will change. In addition, the calculated values for $\lambda$ are consistent with those observed for iron cobalt rhodium molybdenum nitride~\cite{Li16}.

\begin{table}[H]
\setlength{\tabcolsep}{4pt}
\renewcommand*{\arraystretch}{1.3}
\centering
\footnotesize
\caption{Ferromagnetic exchange energy $J$, spin $s$ at 2~K, ferromagnetic exchange stiffness $A$, DM interaction density
         $D$ and expected helical wavelength $\lambda$ for the series of compounds.}
\begin{tabular}{cccccc}
\toprule
 & $J$                  & $s$ & $A$                     & $D$                        & $\lambda$  \\
   $x$    &  (10$^{-21}$ J)&        &  (10$^{-12}$ J\,m$^{-1}$) &(10$^{-4}$ J\,m$^{-2}$)& (nm) \\
\colrule
0    & 2.25 & 0.60 & 3.56 & 1.42 & 157\\
0.25 & 2.37 & 0.55 & 3.21 & 1.68 & 120\\
0.5  & 2.57 & 0.54 & 3.26 & 1.88 & 109\\
0.75 & 2.82 & 0.52 & 3.35 & 2.20 & 96\\
1    & 3.08 & 0.51 & 3.48 & 3.35 & 65\\
\botrule
\end{tabular}
\label{tab:aver_fie}
\end{table}

\begin{figure}[ht]
	\centering
	\includegraphics[height=8cm]{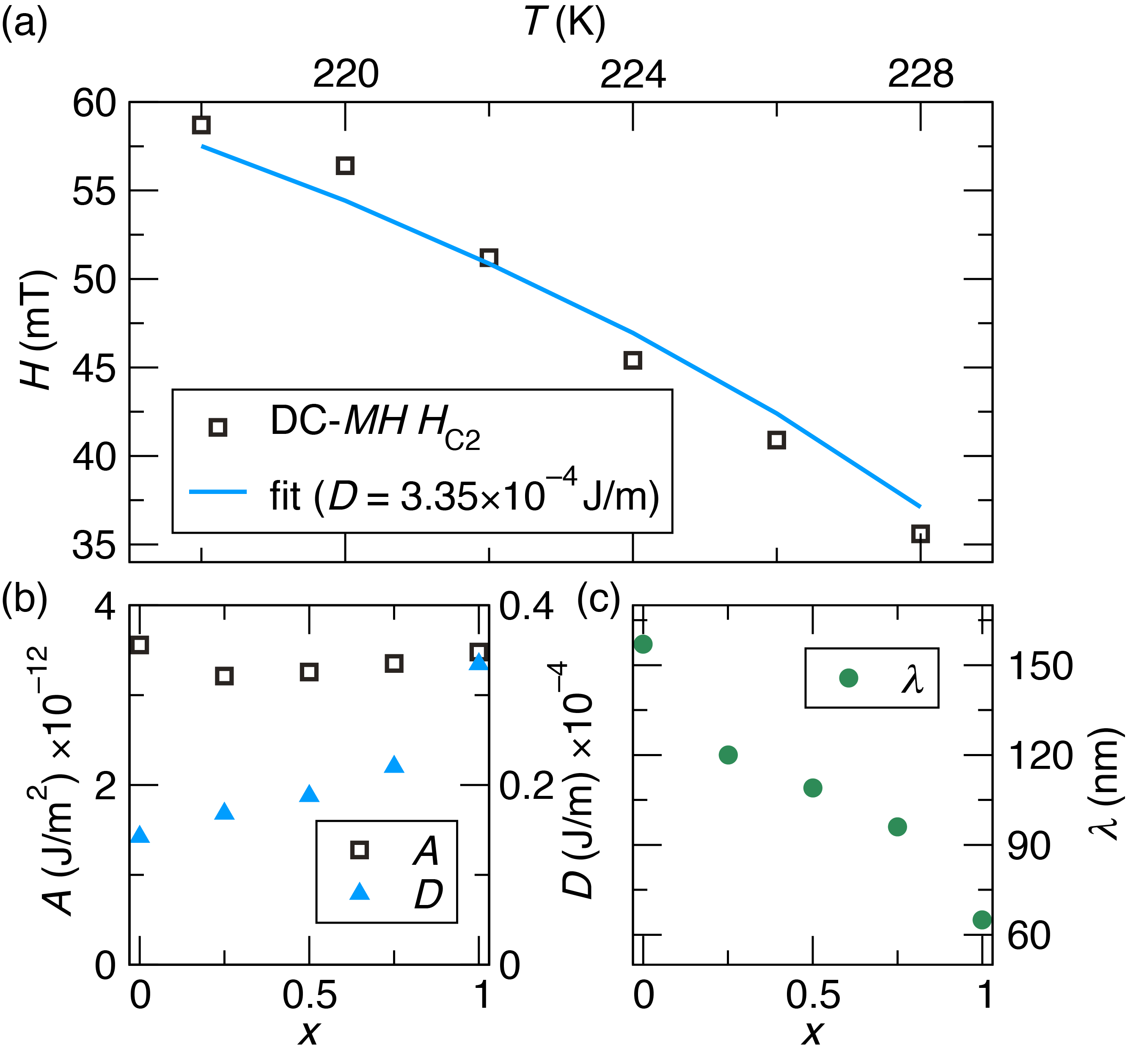}
	\caption{(a) Values of $H_\mathrm{C2}$ for \ce{FePtMo3N} below the Brazovskii transition and the fit 
        used to obtain the parameter $D$ using Eq.~\ref{eq:hc2} and the values for $S(T)$. (b) Exchange stiffness 
        $A$ and DM interaction density $D$ versus the composition of the series $x$. (c) Helical wavelengths $\lambda$ 
        against $x$.}
	\label{fig:a_d_lamb}
\end{figure}

Fig.~\ref{fig:a_d_lamb}~(b) shows the change of $A$ and $D$ with the variation of the composition $x$. The exchange 
stiffness $A$ generally increases slightly with increasing $x$, which corresponds to the increasing Curie temperatures. The 
value for the $x=0$ sample differs from the trend of the series due to the higher $s$ value. Overall, $A$ only changes by 
about 10\,\% across the series. On the other hand, the parameter $D$ is found to increase dramatically with rising $x$, 
increasing by 235\,\% across the series. Therefore, we conclude that the introduction of heavy Pt atoms enhances the 
strength of spin-orbit coupling and therefore DM interaction. This increase is also visible in Fig.~\ref{fig:zfc-mh}~(a), where the larger DM interactions lead to bigger saturation fields and therefore to a smaller susceptibility in the low field magnetic measurements for the compounds with higher Pt content. As a result of the increasing DM interaction strength, the helical period 
length $\lambda$ [Fig.~\ref{fig:a_d_lamb}~(c)] and the skyrmion lattice parameter are expected to decrease from 
157\,nm for $x$\,=\,0 to 65\,nm for $x$\,=\,1.
The increase in spin-orbit coupling also causes the fields needed to 
stabilize the skyrmions to increase from about 50\,mT for $x$\,=\,0 to about 150\,mT for $x$\,=\,1. The proposed skyrmion lattice phases in the polycrystalline nitride samples are with their expected lattice parameters ranging between 50\,nm and 150\,nm suitable for direct observation using small-angle neutron scattering \cite{Gilbert19} or Lorentz transmission electron microscopy \cite{Li16}.

The expected wavelengths $\lambda$ for the compounds in the series FePd$_{1-x}$Pt$_x$Mo$_3$N are comparable to that observed in the near-room-temperature skyrmion host FeGe ($T_C$\,=\,278\,K, $\lambda$\,=\,70\,nm) \cite{Yu2011}. On the other hand, much smaller skyrmions have been observed in MnSi (18\,nm) \cite{Muehlbauer09}; however, this small skyrmion size can be largely attributed to the small ferromagnetic exchange constant, as evidenced by the low $T_C$\,=\,29\,K. On the other hand, Co$_x$Zn$_y$Mn$_z$, which can show a stable skyrmion lattice between at high-temperature (150\,K to 475\,K, depending on composition), shows larger skyrmion lattice parameters of between 115\,nm and 190\,nm \cite{Tokunaga15}. In the room-temperature antiskyrmion host Mn$_{1.4}$PtSn, the lattice parameter ranges from about 150\,nm to 350\,nm depending on temperature and field \cite{Nayak17}. A unique feature in our nitrides is that the skyrmion size can be decreased while simultaneously increasing the Curie temperature. The skyrmion size is reduced by more than 50\% going from \ce{FePdMo3N} to \ce{FePtMo3N}, while the Curie temperature is simultaneously increased by 65\,K. In the compounds Co$_x$Zn$_y$Mn$_z$ ($x+y+z=20$), the change in composition from Co$_9$Zn$_9$Mn$_2$ to Co$_8$Zn$_8$Mn$_4$ decreases the helical period length from 155\,nm to 125\,nm but also decreases $T_C$ from 425\,K to 310\,K. Similar behavior can be seen in the skyrmion hosts Mn$_{1-x}$Ir$_x$Si and Fe$_{1-x}$Mn$_x$Ge, where the decrease of $\lambda$ also leads to significant reduction in $T_C$ \cite{Dhital17,Shibata13}. 

\section{Conclusion}

We have shown that FePd$_{1-x}$Pt$_x$Mo$_3$N compounds in the filled $\beta$-Mn crystal 
structure are promising bulk skyrmion host candidates. High-purity polycrystalline samples with different 
Pd/Pt ratios can be made in a simple gas nitridation procedure, followed by an annealing step under pressurized nitrogen.
DC magnetization and AC susceptibility measurements reveal that these compounds exhibit magnetic phase diagrams consistent with those 
of other chiral cubic skyrmion hosts. In FePtMo$_3$N, an A~phase below 230\,K with 
a width of 8\,K and a height of 10\,mT is seen. The magnetic phase diagrams of these compounds 
are used to determine the ferromagnetic exchange stiffness $A$ and 
Dzyaloshinskii-Moriya interaction density $D$ for each of the compounds. This analysis indicates that $D$ increases 
with increasing Pt content, and, consequently, the skyrmion lattice parameter is predicted to be tunable between 
157\,nm ($x$\,=\,0) and 65\,nm for ($x$\,=\,1). This study demonstrates that the introduction of heavy elements into a 
skyrmion host can be used to control $D$ and achieve small skyrmions; however, for this strategy to be effective 
the substitution must be performed in such a way as to not weaken the ferromagnetic exchange.

\begin{acknowledgments}

This work was supported by Materials Research Science and Engineering Center (MRSEC) program of the National Science 
Foundation through DMR-1720256 (IRG-1). Use of the Advanced Photon Source at Argonne National Laboratory was 
supported by the U. S. Department of Energy, Office of Science, Office of Basic Energy Sciences, under Contract 
No. DE-AC02-06CH11357. J.D.B. is supported by the NSF Graduate Research Fellowship Program under grant no. 1650114.

\end{acknowledgments}

%

\end{document}


\title{Supplemental Material:\\ Controlling Dzyaloshinskii-Moriya interactions in the skyrmion hosts FePd$_{1-x}$Pt$_x$Mo$_3$N}

\author{Linus~Kautzsch}
\affiliation{Materials Research Laboratory, University of California, Santa Barbara, California 93106, USA}
\affiliation{Max Planck Institute for Chemical Physics of Solids, D-01187 Dresden, Germany}

\author{Joshua~D.~Bocarsly}
\affiliation{Materials Research Laboratory, University of California, Santa Barbara, California 93106, USA}
\affiliation{Materials Department, University of California, Santa Barbara, California 93106, USA}

\author{Claudia~Felser}
\affiliation{Max Planck Institute for Chemical Physics of Solids, D-01187 Dresden, Germany}

\author{Stephen~D.~Wilson}
\affiliation{Materials Research Laboratory, University of California, Santa Barbara, California 93106, USA}
\affiliation{Materials Department, University of California, Santa Barbara, California 93106, USA}

\author{Ram~Seshadri}
\affiliation{Materials Research Laboratory, University of California, Santa Barbara, California 93106, USA}
\affiliation{Materials Department, University of California, Santa Barbara, California 93106, USA}
\affiliation{Department of Chemistry and Biochemistry, University of California, Santa Barbara, California 93106, USA}

\date{\today}

\maketitle

\onecolumngrid
\section{Additional tables}

\begin{table}[!hbt]
\setlength{\tabcolsep}{4pt}
\small
\caption{Molar fractions $n_{\rm{i}}$ of the elements in the compounds determined by X-ray fluorescence spectroscopy. Nitrogen was not analyzed, and is ignored in these molar fractions.}
\begin{tabular}{lrrrrr}
\toprule
$x$       & 0          & 0.25       & 0.5        & 0.75       & 1          \\
\colrule
$n_{\rm{Fe}}$   & 0.218   & 0.202   & 0.188   & 0.204   & 0.183  \\
$n_{\rm{Pd}}$   & 0.216   & 0.157   & 0.109   & 0.050   & \multicolumn{1}{c}{--}           \\
$n_{\rm{Pt}}$    & \multicolumn{1}{c}{--}         & 0.055   & 0.104   & 0.158   & 0.202   \\
$n_{\rm{Mo}}$ & 0.566 & 0.586 & 0.599 & 0.588 & 0.615\\
\botrule
\end{tabular}
\label{tab:xrf}
\end{table}

\section{Additional figures}

\begin{figure}[ht]
	\centering
	\includegraphics[height=7cm]{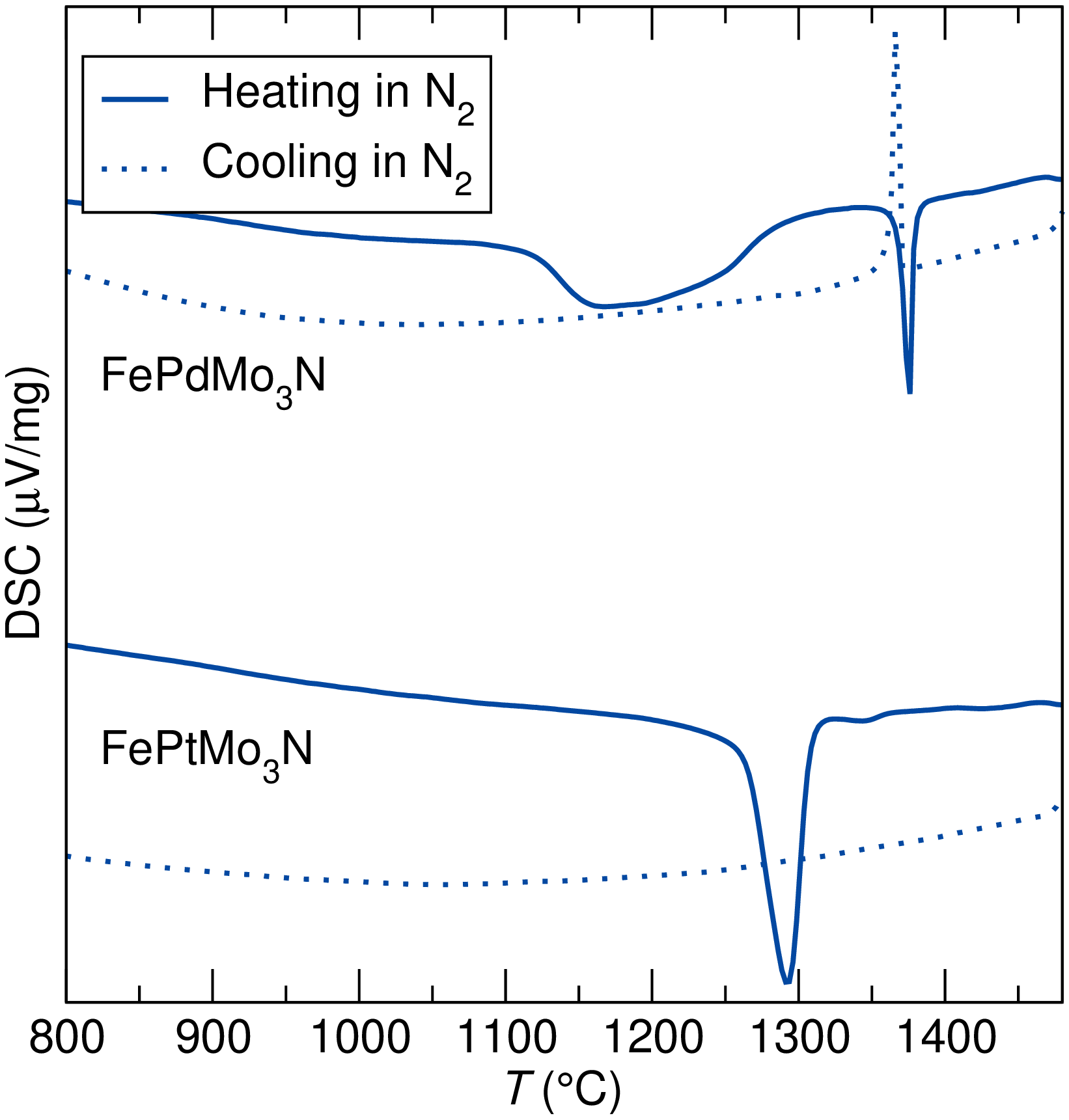}
	\caption{Differential scanning calorimetry for the compounds with $x=0$ and $x=1$ in flowing N$_2$ gas.}
	\label{fig:dsc}
\end{figure}

\begin{figure}[ht]
	\centering
	\includegraphics[height=12cm]{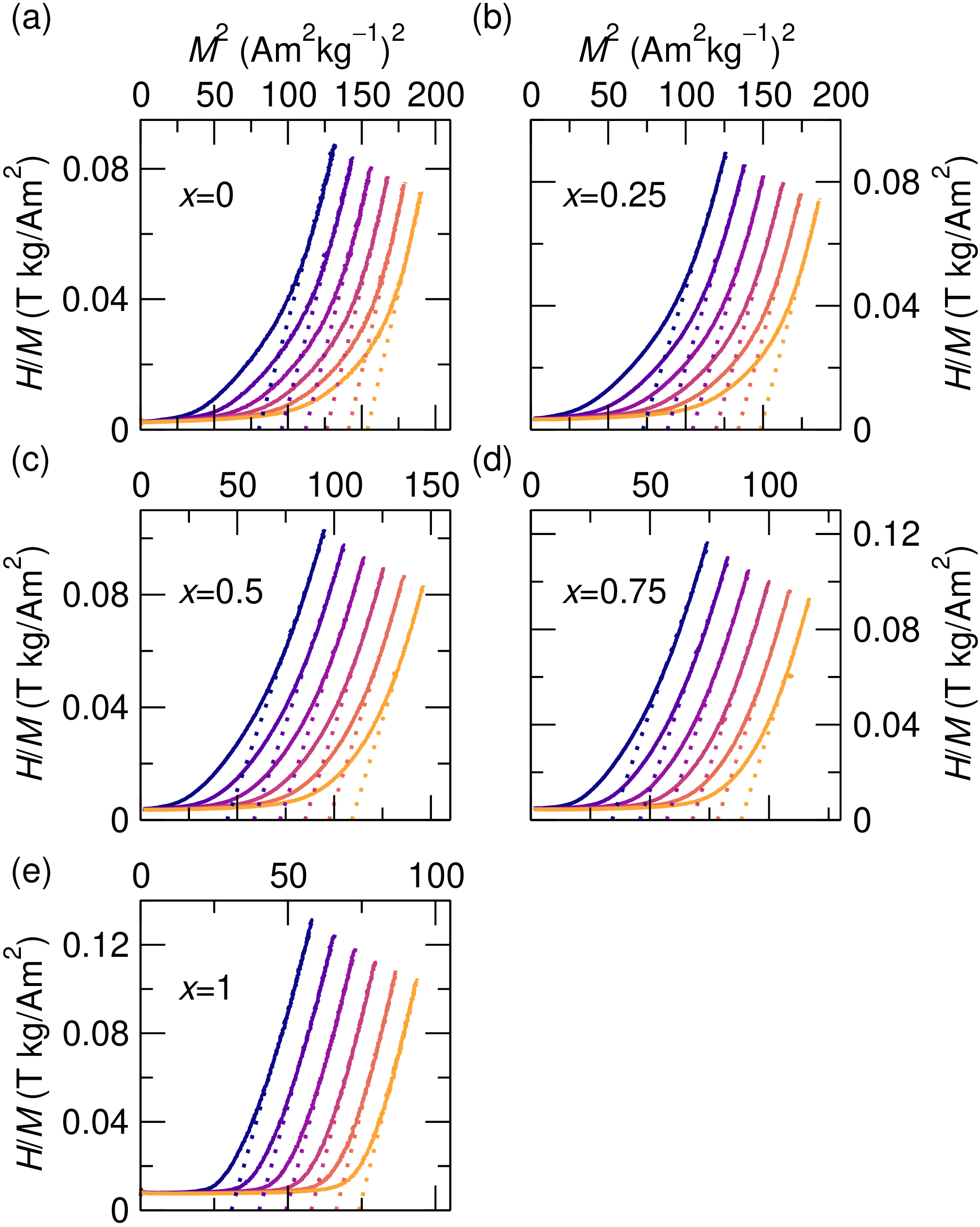}
	\caption{Arrot-Belov plots of magnetization versus magnetic field measurements below the Curie transition for (a) $x=0$, (b) $x=0.25$, (c) $x=0.5$, (d) $x=0.75$ and (e) $x=1$. Linear fits (dotted lines) to the data with values of $H$/$M$ above 0.05\,T\,kg/Am$^2$ have been performed. The $x$ intercepts of the fits correspond to the cubed saturation magnetizations $M_{\mathrm{s}}^2$ at each temperature.}
	\label{fig:arrot}
\end{figure}

\begin{figure}[ht]
	\centering
	\includegraphics[height=8cm]{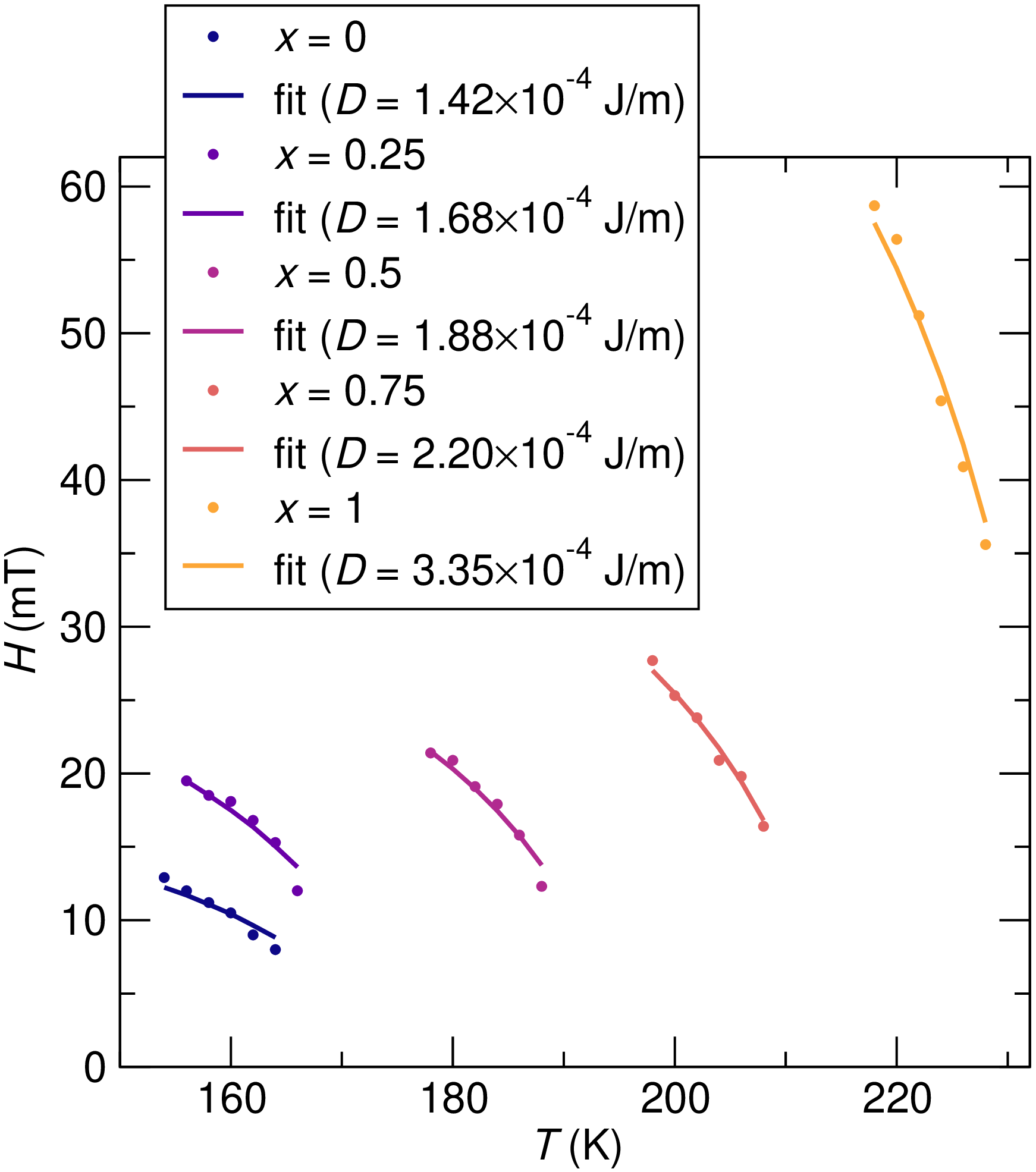}
	\caption{Magnetic fields $H_{\mathrm{C2}}$ (dots) below the fluctuation disordered state from magnetization versus field measurements from Fig.~\ref{fig:arrot} and lines that were fit to the data to obtain $D$.}
	\label{fig:d_fit}
\end{figure}
